\documentclass[aps,pre,amsmath,amssymb,preprint,%linenumbers,
superscriptaddress,
]{revtex4-1}
\ifx\pdfoutput\undefined
\usepackage{graphicx}
\else
\usepackage{graphicx}
\usepackage{epstopdf}
\fi
\usepackage[pdftitle={Yongxiang HUANG},bookmarks=true,citecolor=red,colorlinks=false]{hyperref}
\usepackage{wasysym}
\usepackage{xcolor}

\usepackage{CJK}

\newcommand{\red}[1]{\textcolor{black}{#1}}

\begin{document}

%  \linenumbers
%\setpagewiselinenumbers
%\preprint{APS/PRL}
\begin{CJK*}{GB}{gbsn} % Use default fonts from CJK (see below)
\title{Intermittency Measurement in Two-Dimensional Bacterial Turbulence}
% Force line breaks with %\\

\author{Xiang Qiu (ÇñÏè)}
\affiliation{School of Science, Shanghai Institute of Technology, Shanghai
200235,  China}

\author{Long Ding (¶¡Áú)}
\affiliation{School of Science, Shanghai Institute of Technology, Shanghai
200235,  China}

\author{Yongxiang Huang (»ÆÓÀÏé)}
\email{yongxianghuang@gmail.com}
\affiliation{State Key Laboratory of Marine Environmental Science, College of Ocean and Earth Sciences,
Xiamen University, Xiamen 361102, PR China}

\author{Ming Chen (³ÂÃú)}
%\email{yongxianghuang@gmail.com}
\affiliation{State Key Laboratory of Marine Environmental Science, College of Ocean and Earth Sciences,
Xiamen University, Xiamen 361102, PR China}

\author{Zhiming Lu (¬־Ã÷)}%
%\email{francois.schmitt@univ-lille1.fr}
\affiliation{Shanghai Institute of Applied Mathematics and Mechanics, Shanghai Key Laboratory of Mechanics in Energy Engineering, Shanghai University,
Shanghai 200072,  China}

\author{Yulu Liu (ÁõÓî½)}%
%\email{francois.schmitt@univ-lille1.fr}
\affiliation{Shanghai Institute of Applied Mathematics and Mechanics, Shanghai Key Laboratory of Mechanics in Energy Engineering, Shanghai University,
Shanghai 200072,  China}

\author{Quan Zhou (ÖÜÈ«)}%
%\email{francois.schmitt@univ-lille1.fr}
\affiliation{Shanghai Institute of Applied Mathematics and Mechanics, Shanghai Key Laboratory of Mechanics in Energy Engineering, Shanghai University,
Shanghai 200072,  China}

\date{\today}% It is always \today, today,
             %  but any date may be explicitly specified

\begin{abstract}
In this paper, an experimental velocity database of a bacterial collective motion , e.g., \textit{B. subtilis}, in turbulent phase with volume filling fraction $84\%$  provided by Professor Goldstein at the Cambridge University UK,  was analyzed to emphasize the scaling behavior of this active  turbulence system. 
This was accomplished by performing a Hilbert-based methodology analysis
to retrieve the scaling property without the $\beta-$limitation. A dual-power-law behavior  separated by the viscosity scale $\ell_{\nu}$ was observed  for the $q$th-order Hilbert moment $\mathcal{L}_q(k)$.  This dual-power-law belongs to an inverse-cascade since the scaling range is above the injection scale $R$, e.g., the bacterial body length.  The measured scaling exponents $\zeta(q)$ of both the small-scale \red{(resp. $k>k_{\nu}$) and large-scale (resp. $k<k_{\nu}$)} motions are convex, showing the multifractality. A lognormal formula was put forward to characterize the multifractal intensity. The measured intermittency parameters are  $\mu_S=0.26$ and $\mu_L=0.17$ respectively for the small- and large-scale motions. It implies that  the former cascade is more intermittent than the latter one, which is also confirmed by   the corresponding singularity spectrum $f(\alpha)$ vs $\alpha$.  Comparison with the conventional two-dimensional Ekman-Navier-Stokes equation, a continuum model indicates that the origin of the multifractality could  be a result of some additional nonlinear  interaction terms, which deservers a more careful investigation. 
\end{abstract}

%\pacs{47.27.eb,94.05.Lk, 47.27.Gs}%{Time series analysis}
%\pacs{02.50.Fz}{Stochastic analysis}
%\pacs{47.27.Gs}{Isotropic turbulence; homogeneous
%turbulence}
%\pacs{47.53.+n}{Fractals in fluid dynamics}
% PACS, the Physics and Astronomy
                             % Classification Scheme.
%\keywords{autocorrelation function, power law}%Use showkeys class option if keyword
                              %display desired
\maketitle
\end{CJK*}
\section{Introduction}

\red{The most fascinating  aspect of the hydrodynamic turbulence is its  scale invariance, which is conventionally characterized by the $q$th-order structure functions, 
\begin{equation}
S_q(\ell)=\langle \vert \Delta \mathbf{u}_{\ell}(\mathbf{x},t) \vert ^q \rangle_{\mathbf{x},t}\sim \ell^{\zeta(q)}\label{eq:SF}
\end{equation}
 where $\Delta \mathbf{u}_{\ell}(\mathbf{x},t)=\mathbf{u}(\mathbf{x}+\ell,t)-\mathbf{u}(\mathbf{x},t)$ is velocity increment of the Eulerian velocity field,  $\ell$ is the separation scale, and $\langle \,\cdot\,\rangle_{\mathbf{x},t}$ means an ensemble average over $\mathbf{x}$ and $t$
 \citep{Frisch1995}. The separation scale $\ell$ should lie in the so-called inertial range $\ell_{\nu}\ll\ell\ll L$, where $\ell_{\nu}$ is known as the Kolmogorov scale or viscosity scale, and $L$ is the integral length scale.  It was first introduced by \citet{Kolmogorov1941} in the year 1941 (resp. K41 for short) with a non-intermittent scaling exponent  $\zeta(q)=q/3$ \citep{Frisch1995}. The K41 theory is deeply related with an idea of  energy cascade, which was first introduced phenomenologically 
  by Richardson in the year 1922 \citep{Richardson1922}.  The energy cascade has been interpreted as a main feature of the energy conservation law in the 3D turbulence, in which the energy is transferred from   large-scale structures to small-scale ones, until the viscosity scale $\ell_{\nu}$, where the kinetic energy is converted into heat \citep{Frisch1995}. 
   Generally for a mono-fractal process, for instance fractional Brownian motion, a self-similarity process with stationary increments on different separation scales $\ell$, the scaling $\zeta(q)$ is linear with $q$, e.g., $\zeta(q)=qH$, where $H$ is the so-called Hurst number. 
  However,  for the high-Reynolds number turbulent flows, the experimental $\zeta(q)$ obtained from various experiments  and numerics deviates from the K41 value $q/3$ \citep{Anselmet1984,Sreenivasan1997,Warhaft2000,Lohse2010}. 
A concept of  multifractality/multiscaling  is  put forward to interpret this  deviation \cite{Benzi1984,Parisi1985}. 
It is further recognized  as a main result of  the energy dissipation field intermittency
 \citep{Frisch1995}. The  `intermittent' or `intermittency' of the small-scale fluctuation was firstly noticed experimentally  by \citet{Batchelor1949}.  It  means a huge small-scale variation  of the  energy dissipation rate, see a nice example in Ref. \cite[see Fig.\,1]{Meneveau1991JFM} or in Ref. \cite[see Fig.\,2.3]{Schmitt2016Book}.  It is  a result of strong nonlinear interactions in the Navier-Stokes equations.  Several theoretical models have been put forward to describe the intermittent property of the energy dissipation field, for instance, the lognormal model \cite{Kolmogorov1962}, log-Poisson model \cite{She1994PRL,Dubrulle1994}, log-stable model \cite{Schertzer1987,Kida1991}, to list a few.   Multifractality has also been  recognized as a common feature of complex dynamic systems, such as financial activities \citep{Mantegna1996,Schmitt1999,Li2014PhysicaA}, wind energy \citep{Calif2013PHYSICA}, geosciences \citep{Huang2009Hydrol,Schmitt2009JMS}, to name a few.
}

 In the 2D turbulence,  an additional  enstrophy (i.e. the square of vorticity $\Omega=\frac{1}{2}\omega^2$)  conservation  is emerging  below the forcing scale $\ell_F$ as a forward enstrophy cascade. On the other hand, above this forcing scale, the energy conservation leads to  an inverse energy cascade, forming a remarkable large-scale motion, which could reach the  system size  \citep{Xia2011NatPhys}. Note that both the energy and enstrophy are injected into the system via the forcing scale $\ell_F$. A 2D turbulence  theory has been put forward in the year 1967 by \citet{Kraichnan1967PoF} to interpret this dual-cascade phenomenon. This 2D turbulence theory has been recognized as ``one of the most
important results in turbulence since Kolmogorov's 1941 work" \citep{Falkovich2006PhysToday}.
More precisely, there is a forward enstrophy cascade with $E(k)\sim k^{-3}$ when $k_F\ll k\ll k_{\nu}$, in which $k_{F}$ is the forcing \red{wavenumber}, and $k_{\nu}$ is the viscosity \red{wavenumber} where the enstrophy is dissipated; and there is an inverse cascade with $E(k)\sim k^{-5/3}$ when $k_{\alpha}\ll k\ll k_{F}$, in which $k_{\alpha}$ is the Ekman friction \red{wavenumber} \citep{Kraichnan1967PoF,Boffetta2012ARFM}. 
This 2D turbulence theory has been partially confirmed by experiments and numerical simulations for the velocity field \citep{Boffetta2012ARFM}. However, the statistics of the vorticity field shows inconsistence \citep{Paret1999PRL,Kellay1998PRL,Tan2014PoF}.
Concerning the multifractality, an extremely important feature of the turbulent systems, the inverse energy cascade is non-intermittent or anomaly-free, which was confirmed by experiments not only using the velocity field \citep{Falkovich2006PhysToday,Wang2015JSTAT}, but also the vorticity field \citep{Tan2014PoF}. 
However, on the other hand, it has long been controversial  whether or not the forward enstrophy cascade is intermittent   since the classical structure function analysis fails to detect the scaling behavior when the slope of the Fourier power spectrum is $\beta\ge 3$ \citep{Frisch1995,Huang2010PRE}.  \citet{Nam2000PRL} theoretically  showed   that  when the Ekman friction is present, the forward enstrophy cascade is then intermittent \citep{Bernard2000EPL}. As already mentioned above   this result is difficult to verify experimentally by using the conventional structure function analysis since the convergence condition requires the scaling exponent $\beta$ of the Fourier spectrum, i.e., $E(k)\sim k^{-\beta}$, to be in the range $(1,3)$ \cite{Frisch1995,Huang2010PRE,Schmitt2016Book}, see also discussion in Sec.\,\ref{sec:SF}. This is known as the $\beta-$limitation. Recently, Tan, Huang \& Meng \citep{Tan2014PoF} applied the Hilbert-Huang transform, a method free with $\beta$-limitation,  to the vorticity field obtained from a high-resolution numerical simulation database with resolution $8192^2$ grid points. They confirmed that the forward enstrophy cascade is intermittent, and the inverse cascade is non-intermittent. Wang \& Huang \citep{Wang2015JSTAT}
proposed a   $\beta-$limitation free multi-level segment analysis and applied it to the 2D velocity field. They confirmed again that the forward enstrophy cascade is   intermittent when considering the velocity statistics .

Specifically for a bacterial suspension in a thin fluid, \red{if the considered spatial size is much larger than the thickness of the suspension,} it could be approximated as a 2D fluid system.   In a such system,
the fluid is stirred by the bacterial activities at their body length $R$. Due to the hydrodynamic interaction or other mechanisms, the flow  exhibits a 
 turbulent-like movement, showing multiscale statistics \citep{Wu2000PRL,Pooley2007PRL,Ishikawa2008PRL,Rushkin2010PRL,Ishikawa2011PRL,
Chen2012PRL,Wensink2012PNAS,Dunkel2013PRL,Saintillan2012JRSI,Dunkel2013NJP,Grobmann2014PRL,Marchetti2013RMP}.
  Such flows are then called as bacterial turbulence or active turbulence.
In this special flow system, the energy is injected into the system via the scale of the bacterial body length $R$ typically around few $\mu$m \citep{Wensink2012PNAS}. The flow velocity is also of the order of few $\mu$m per second. The corresponding Reynolds number is about $Re=\mathcal{O}(10^{-3})$.  In the traditional view of the classical hydrodynamic   turbulence, the flow at such low Reynolds number is laminar without turbulent-like statistics.   It is surprising that the statistics of the active fluid exhibits a turbulent-like fluctuation, e.g.,  long range correlation of velocity
\citep{Dombrowski2004,Ishikawa2008PRL,Ishikawa2011PRL,Chen2012PRL,Dunkel2013PRL,Grobmann2014PRL}, power-law behavior \citep{Grobmann2014PRL,Saintillan2012JRSI,Wu2000PRL,Wensink2012PNAS,Liu2012PRE}, etc.
For example,  Wu \& Libchaber  reported that due to the collective dynamics of bacteria in a freely suspended soap film, the measured mean displacement function of  beads demonstrates a superdiffusion in short times and normal diffusion in long times \citep{Wu2000PRL}. 
Wensink \textit{et al., } \citep{Wensink2012PNAS}
observed  a dual-power-law (DPL)   behavior in a quasi-2D active fluid.
Due to the viscosity damping by the low-Re solvent, the experimental power-law behavior extends roughly up to \red{$\ell_{\nu}\simeq10
R\simeq 50\,\mu$m}, corresponding to a wavenumber $k_{\nu}/k_{R}\simeq0.1$, where $k_{R}=1/R$ is the wavenumber of the bacterial body length, and $\ell_{\nu}$ is the viscosity scale  \footnote{The Kolmogorov scale or viscosity scale is estimated as $\ell_{\nu}=(\nu^3/\epsilon)^{3/4}$, in which $\nu$ is the viscosity of the fluid and $\epsilon$ is the energy dissipation rate. A typical value of $\ell_{\nu}$ in the ocean is $0.3\sim 2\,$mm. A typical $\ell_{\nu}$ in a pipe flow is around $25\,\mu$m with a diameter $50\,$mm and a velocity $1.8\,$m$/$s. For the current database, it is reasonable to take the viscosity scale as $\ell_{\nu}\simeq 50\,\mu$m.}. Above this wavenumber, e.g., $0.1\le k/k_{R}\le 1$, one may has the energy-inertial regime of classical turbulence with a power-law roughly  as $E(k)\sim k^{-8/3}$; and  below it, e.g., $k/k_{R}\le 0.1$,  but not far from the viscosity scale $k_\nu$,  the viscous damping  play an important role  with a power-law that roughly can be fitted as $E(k)\sim k^{5/3}$ \citep{Wensink2012PNAS}. 
It is worth to point out here that these two power-laws are on the same side of the injection scale $R$. Both of them    belong to the inverse cascade.
To the best of our knowledge, there are  very few works related with the multifractality of the bacterial turbulence since the  structure function analysis fails to capture the scaling behavior.    Liu \& I
\citep{Liu2012PRE} experimentally found that the multifractality revealed by the extended self-similarity (ESS) technique is increasing with the cell concentration.   Note that in the ESS approach, instead of plotting  the $q$th-order structure function $S_q(\ell)$  versus the separation scale $\ell$, the experimental  $S_q(\ell)$ is often plotted against with  $S_2(\ell)$ or   $S_3(\ell)$ \cite{Benzi1993PRE}.  It provides a more robust way to extract the scaling exponent $\zeta(q)$ \cite{Benzi1995,Benzi1993EPL,Arneodo1996}. With the help of ESS, the relative scaling exponent is found to be universal for a large range of Reynolds number and the statistics order $q$ up to 10 \cite{Benzi1995}.

 In this paper, we  investigated the multifractality of the bacterial turbulence experimentally using the Hilbert-Huang transform  to identify the power-law behavior and extract scaling exponent $\zeta(q)$ directly without resorting to the ESS technique.
It is found that the intermittent correction is relevant in the observed DPL. The corresponding intermittency parameter provided by a lognormal formula is $\mu_S=0.26$ and $\mu_L=0.17$ respectively for the small-scale fluctuations above the viscosity scales and the large-scale fluctuations below the viscosity scales.  The observed multifractality could be a result of the several additional nonlinear terms appearing in an Ekman-Navier-Stokes-like model equation \citep{Wensink2012PNAS}.

\section{Experimental data}

\begin{figure}
\centering
\includegraphics[width=0.95\linewidth,clip]{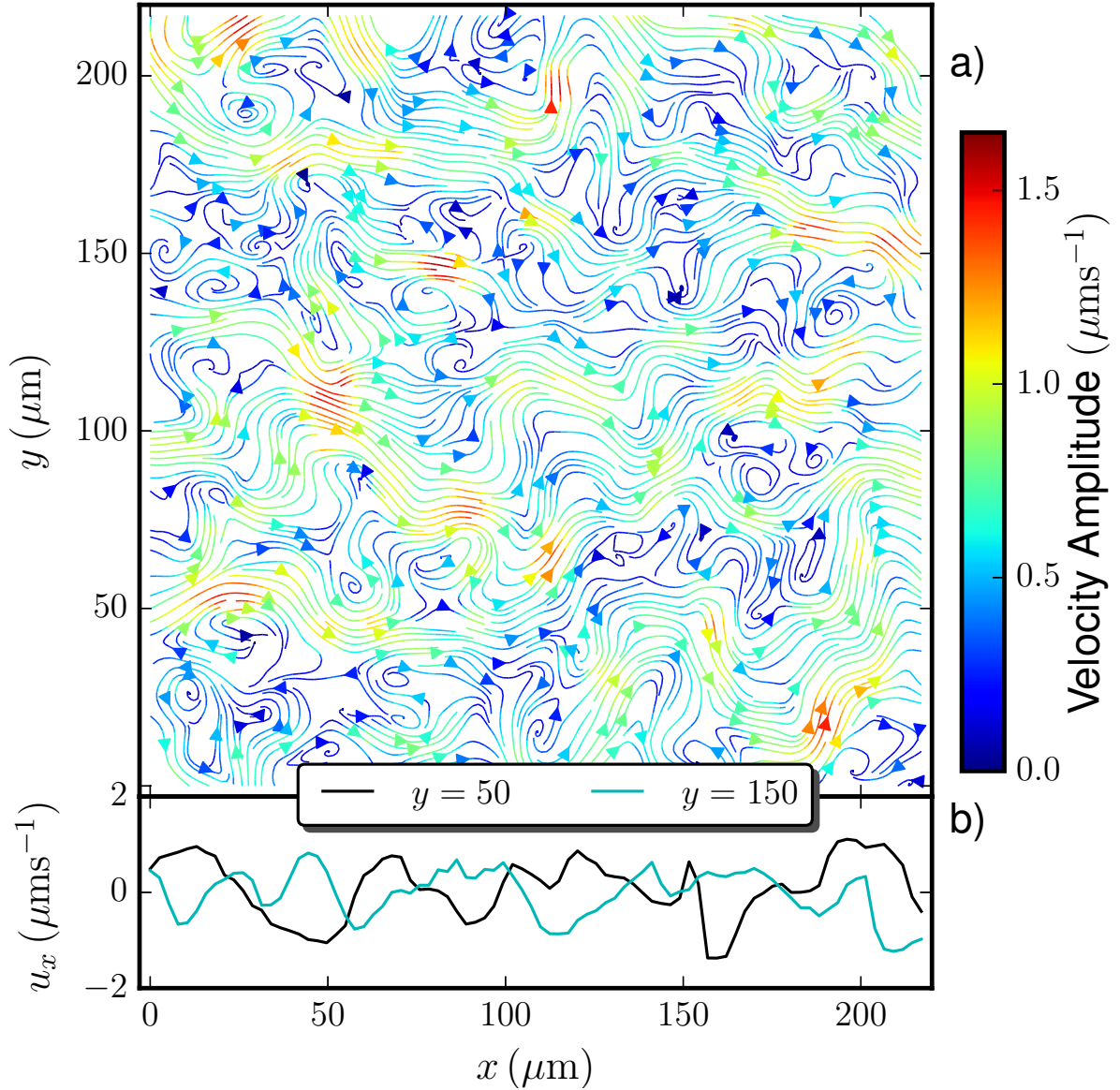}
 \caption{(Color online) %A snapshot of the velocity field $\mathbf{u}(x,y)$. The velocity vector  is color coded by the amplitude of the velocity.  The color background is the corresponding  energy.
\red{a) A snapshot of the velocity streamline, which is color encoded by the velocity amplitude. b) The velocity $u_x(x)$ at $y=50\,\mu\mathrm{m}$ and $y=150\,\mu\mathrm{m}$.
Energetic structures are observed roughly with a spatial scale $\sim50\,\mu$m, corresponding to $10$ times of the bacterial body size $R$.}  }
 \label{fig:snap}
 \end{figure}

The experiment data analyzed here is provided by  Professor R.E. Goldstein at the  Cambridge University UK. We recall  briefly the main parameters of this quasi-2D experiment in a microfluidic chamber.  The bacteria used in this experiment is \textit{B. subtilis} with an individual  body length approximately $5\,\mu$m, in which the energy is injected into the system. The volume filling fraction is $\phi=84\%$ with bacterial number $N\simeq9968$ and aspect ratio $a=5$, i.e., the ratio between the bacterial body length $R$ and the body diameter. The quasi-2D microfluidic chamber is with a vertical height $H_c$ less or equal to the individual body length of \textit{B. subtilis} (approximately 5$\,\mu$m). With these  parameters, the flow is then in a turbulent phase \citep{Wensink2012PNAS}. The PIV (particle image velocimetry) measurement area is $217\mu\mathrm{m}\times 217\mu\mathrm{m}$.  The image resolution is of $700\,$pix$\times700\,$pix with conversion rate $0.31\,\mu$m/pix and frame rate  $40$Hz. The commercial PIV software Dantec Flow Manager is used to extract the flow field component with a moving window size $32\,$pix$\times 32\,$pix and $75\%$ overlap. This results a  $84\times 84$ velocity vector and a total 1015 snapshots,  corresponding to a time period $\sim25\,$seconds.
Therefore, totally we have $7,161,840$ data points, which ensures a good statistics at least up to the statistical order $q=4$. 

\red{Figure \ref{fig:snap}\,a) shows a snapshot of the streamline, where the velocity amplitude is encoded in color. Figure \ref{fig:snap}\,b) shows the velocity $u_x(x\vert y)$ slice  at $y=50$ and $150\,\mathrm{\mu m}$.  }
% velocity field $\mathbf{u}(x,y)$, where the velocity vector is color coded by of the velocity amplitude. For comparison, the instantaneous energy is also shown as background. 
Visually, we observe energetic structures roughly with  a spatial scale $\sim 50\,\mu$m, corresponding to ten times of the bacterial body size, i.e., $10R$. The origin of this structure is unclear. We will turn back to this point in Sec.\,\ref{sec:discussion}. The flow field is homogeneous and isotropic.  In the following analysis, only the velocity component $u_x(x,y,t)$ is considered. It is first divided into 84 lines along the direction $x$. Statistical quantities are then estimated for all snapshots. 

\section{Scale Mixture Problem of Structure Function Analysis}\label{sec:SF}
\begin{figure}
\centering
\includegraphics[width=0.95\linewidth,clip]{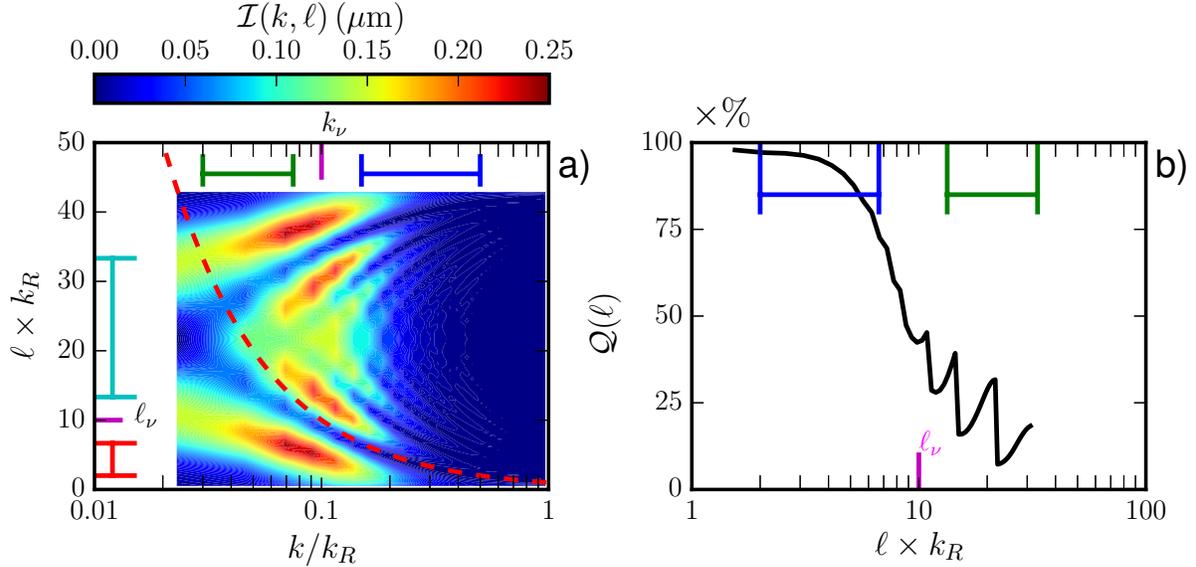}
 \caption{(Color online) a)  The experimental contribution kernel $\mathcal{I}(k,\ell)$, in which the range of the dual-power-law is indicated by horizontal lines. The peak location (resp. the viscosity scale) of the Fourier power spectrum $E(k)$ is indicated by $k_{\nu}$, corresponding to a spatial scale $\ell_{\nu}=1/k_{\nu}$. The dashed line illustrates $\ell=1/k$. b) The measured large-scale contribution $\mathcal{Q}(\ell)=\int_0^{1/\ell} \mathcal{I}(k,\ell)d k$.  }
 \label{fig:Contribution}
 \end{figure}

We  show  here the scale mixture problem of the conventional structure function analysis. The second-order structure function $S_2(\ell)$ can be associated  with the Fourier power spectrum $E(k)$ via the Wiener-Khinchin theorem \cite{Frisch1995,Schmitt2016Book},  
\begin{equation}
S_2(\ell)=\int_0^{+\infty}E(k)\left( 1-\cos(2\pi k \ell) \right) d k
\end{equation} where $\ell$ is the separation scale, $k$ is the wavenumber. A prefactor is ignored. It implies that except for the case $k=n/\ell$, $n=0,1,2,\cdots$, all Fourier components have contribution to $S_2(\ell)$.  Or in other words,  it contains informations  from different Fourier components \citep{Huang2010PRE}.  Taking a pure power law form $E(k)\sim k^{-\beta}$, the convergence conditions at $k\rightarrow 0$ and $k\rightarrow +\infty$ require $\beta\in (1,3)$ \cite{Frisch1995,Huang2010PRE,Schmitt2016Book}. Unfortunately, if the data set has energetic structures,  the structure function analysis will be strongly biased. For instance,  the ramp-cliff structure in the passive scalar turbulence \cite{Huang2010PRE,Huang2011PRE}, vortex trapping event in the Lagrangian velocity \cite{Huang2013PRE}, high intensity vortex in 2D turbulence \cite{Tan2014PoF,Wang2015JSTAT}, daily cycle or annual cycle in the collected geosciences data \cite{Huang2009Hydrol}, to list a few. Therefore, before applying the structure function analysis, as we will show below, it is better to perform a scale-by-scale analysis to see whether such influence exists or not.
To characterize quantitatively  the relative contribution of different Fourier components, we introduced here a contribution kernel function $\mathcal{I}(k,\ell)$, 
\begin{equation}
\mathcal{I}(k,\ell)=\frac{E(k)\left( 1-\cos(2\pi k \ell) \right)}{S_2(\ell)}
\end{equation}
where $E(k)$ is the Fourier power spectrum provided the experimental velocity field. 
Figure \ref{fig:Contribution}\,a) shows the experimental $\mathcal{I}(k,\ell)$, in which the power-law range $0.03<k/k_R<0.075$ and $0.15<k/k_R<0.5$  (see  analysis result in Sec.\,\ref{sec:result}) are illustrated by  solid lines. The dashed line indicates $\ell=1/k$. Visually, most of the contribution is coming from the large-scale part, i.e., $ k/k_R\le 0.2$. It also displays an up-down symmetry.  
This is because the Fourier power spectrum $E(k)$ increasing with $k$ when $k/k_R\le 0.1$ \red{and taking its peak at $k/k_R\simeq 0.1$}, see Figure \ref{fig:spectrum}\,a).  A relative cumulative function is introduced to characterize the relative contribution from the large-scale part, 
\begin{equation}
\mathcal{Q}(\ell)=\int_0^{1/\ell}\mathcal{I}(k,\ell) d k\times 100\%
\end{equation}
Figure \ref{fig:Contribution}\,b) shows the measured $\mathcal{Q}(\ell)$, in which the expected power-law range is indicated by solid line.  Experimentally, $S_2(\ell)$ in the first power-law range, i.e., $0.15< k/k_R< 0.5$, is strongly influenced by the large-scale motions;  in the second power-law range, i.e., $0.03<k/k_R<0.075$, it is strongly influenced by the energetic structures around $k/k_R\simeq 0.1$.  Due to the presence of  energetic structures, the expected  power-law behavior is then destroyed or biased in the physical domain \citep{Wensink2012PNAS}.  A similar phenomenon   has been observed for the vorticity field of the traditional 2D turbulence  with high intensity vortex structures \citep{Tan2014PoF}, and for passive scalar turbulence with ramp-cliff structures  \citep{Huang2010PRE}, etc. 
More details about this topic, we refer the readers to  Ref. \citep{Schmitt2016Book}.

\section{Hilbert-Huang Transform}
 \begin{figure}
\centering
\includegraphics[width=0.95\linewidth,clip]{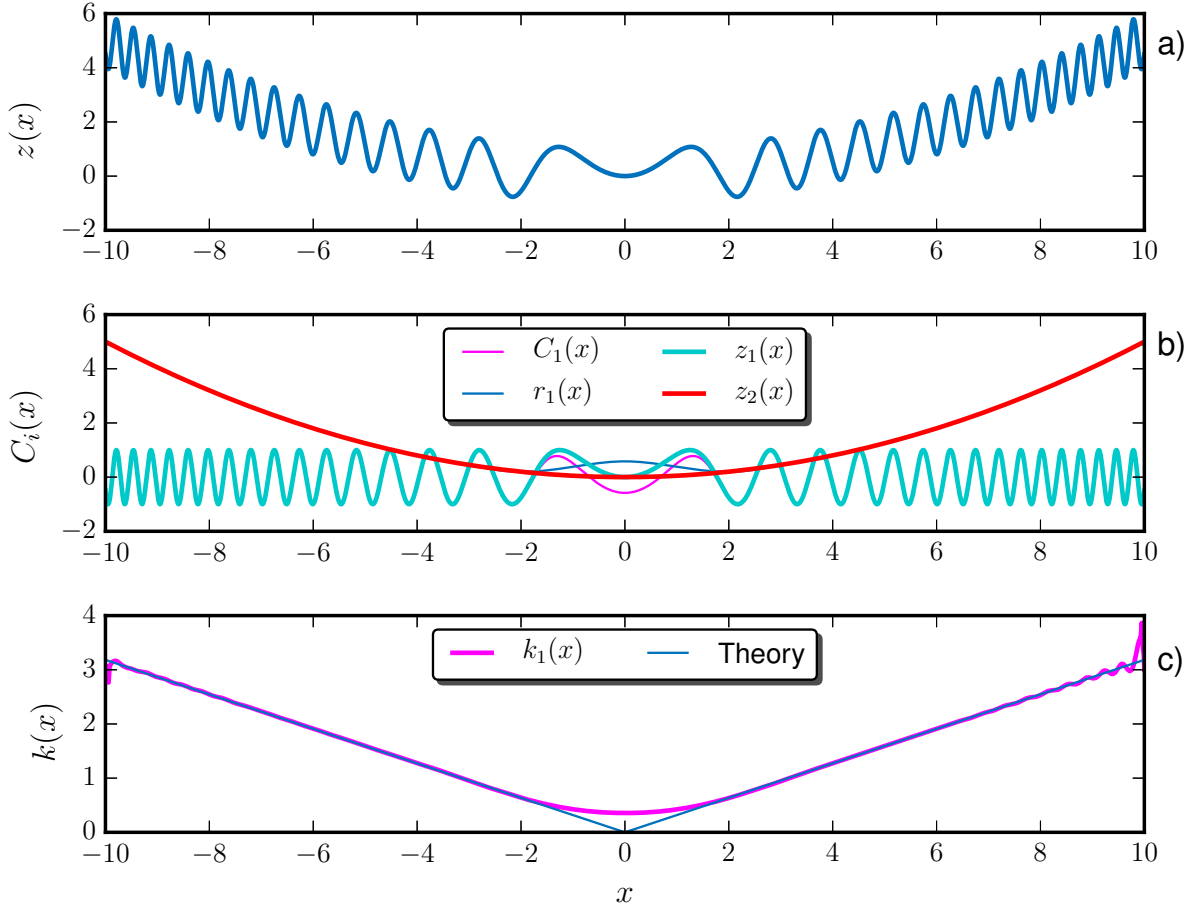}
 \caption{(Color online) a)  Illustration of a toy model $z(x)=z_1(x)+z_2(x)$, where $z_1(x)=\sin(x^2)$, and $z_2(x)=x^2/20$.  b) IMF mode  
 $C_1(x)$ and residual $r_1(x)$
 obtained from EMD algorithm (thin lines). For comparison, the $z_1(x)$ and $z_2(x)$ are also shown (thick lines). c) The measured instantaneous wavenumber $k(x)$ for $C_1(x)$ (thick line), where the theoretical value is shown as a thin line. }
 \label{fig:example}
 \end{figure}
 
In this work, we will employ  a $\beta$-limitation free approach, namely Hilbert-Huang transform \citep{Huang1998EMD,Huang2008EPL}.  It has the capability to isolate different events not only in the physical domain, but also in spectral space 
 \citep{Huang2011PRE,Huang2013PRE,Tan2014PoF,Schmitt2016Book}. This method consists two steps:  \romannumeral1) Empirical Mode Decomposition (EMD), and  \romannumeral2)
 Hilbert spectral analysis. In the following, we present more details of this Hilbert-based approach.

\subsection{Empirical Mode Decomposition}

In reality, most of the collected signals are multi-component, which means that different time or space scales are coexistent \citep{Huang1998EMD,Huang1999EMD}. It is thus necessary to apply a proper method to separate a given signal into a sum of mono-components to have a better view of them.  For example,
in the classical Fourier analysis, a trigonometric function sine or cosine is chosen as the mono-component \citep{Cohen1995}.  
The given data set is then associated with the energy (the square of the amplitude) and the wavenumber (the inverse of the period of the given sine or cosine wave), known as the Fourier power spectrum.

In this Hilbert-based approach,
the so-called  Intrinsic Mode Function (IMF) has been put forward to represent the mono-component, which satisfies the following two conditions: (\romannumeral1)
 the difference between
the number of local extrema and the number of zero-crossings must be zero or one; (\romannumeral2) the running mean value
of the envelope
defined by the local maxima and the envelope defined by the local minima is zero  \citep{Huang1998EMD,Rilling2003EMD}.  Each IMF then has a well-defined Hilbert spectrum \citep{Huang1998EMD}. It allows both the  amplitude- and frequency/wavenumber-modulation  simultaneously since its characteristic scale is defined as the distance between two successive extreme points \citep{Huang2014JSTAT}.

The  Empirical Mode Decomposition algorithm is put forward to extract the IMF modes from a given data set,  e.g., velocity
 $u(x)$.
 The first step of the EMD algorithm is to identify all the local maxima (resp. minima) points. Once all the local  maxima points
are identified,   the upper  envelope $e_{\max}(x)$ (resp. lower envelope $e_{\min}(x)$) is constructed by  a cubic spline interpolation \citep{Huang1998EMD,Huang1999EMD,Flandrin2004EMDa}.  Note that other approaches are also possible to construct the envelope \citep{Chen2006ACM}.   The running mean between these two envelopes
is defined as, 
\begin{equation}
 m_{1}(x)=\frac{(e_{\max}(x)+e_{\min}(x))}{2},
 \end{equation}
 The first component is estimated as, 
 \begin{equation}
h_1(x)=u(x)-m_{1}(x),
\end{equation}
 Ideally,  $h_1(x)$ should
be an IMF as expected.  In reality, however,  $h_1(x)$  may  not satisfy
the condition to be an IMF. We take $h_1(x)$ as a new data series and repeat the sifting process $j$ times, until
$h_{1j}(x)$ is an IMF.
We thus have the first IMF component, 
\begin{equation}
C_{1}(x)=h_{1j}(x),
\end{equation}
and the residual, 
\begin{equation}
r_{1}(x)=u(x)-C_{1}(x),
\end{equation}
%from the data $x(t)$.
 The sifting procedure is then repeated
on residuals  until  $r_n(x)$ becomes a monotonic function or at most has one
local extreme point. This means  that
 no more IMF can be extracted  from $r_n(x)$.  Thus, with this algorithm we finally have $n$
IMF modes with one residual $r_n(x)$. The original data $u(x)$ is then rewritten  as, 
\begin{equation}
u(x)=\sum_{i=1}^{n}C_i(x)+r_{n}(x)
\end{equation}
A  stopping criterion has to be introduced in the EMD algorithm to stop
the sifting process \citep{Huang1998EMD,Huang1999EMD,Rilling2003EMD,Huang2003EMD}.
The first stopping
 criterion is a  Cauchy-type convergence criterion proposed by \citet{Huang1998EMD}. A standard
deviation  defined for two successive sifting processes is written as,
\begin{equation}
\mathrm{SD}=\frac{\sum_{x=0}^{L}\vert
h_{i(j-1)}(x)-h_{j}(x)\vert^2}{\sum_{x=0}^{L} h_{i(j-1)}^2(x)}
\end{equation}
in which $L$ is the total length of the data.
If a calculated SD is smaller than a given value, then the sifting stops,
and gives an IMF.
A typical value   SD$\in[0.2, 0.3]$ has been proposed based on Huang et al.\rq{}s experiences \citep{Huang1998EMD,Huang1999EMD}.
Another widely used criterion is based on three thresholds $\alpha$, $\theta_1$, and
$\theta_2$, which are designed  to guarantee globally small fluctuations 
  meanwhile taking into account locally large
excursions  \citep{Rilling2003EMD}.
The mode amplitude and evaluation function are  given as,
\begin{equation}
a(x)=\frac{e_{\max}(x)-e_{\min}(x)}{2},\,\, \sigma(x)=\vert m(x)/a(x)\vert
\end{equation}
so that the sifting is iterated until $\sigma(x)<\theta_1$ for some prescribed
fraction $1-\alpha$ of the total duration, while $\sigma(x)<\theta_2$ for
the remaining fraction. Typical values proposed in Ref.\,\citep{Rilling2003EMD}  are
$\alpha \approx 0.05$, $\theta_1
\approx 0.05$ and $\theta_2 \approx 10 \,\theta_1$, respectively based on
their experience.
In practice, a maximal iteration number (e.g., $300$) is also chosen to avoid over-decomposing the data set.

A main drawback  of this method is that EMD is an algorithm in practice without rigorous mathematical foundation \cite{Huang1998EMD}. Several works attempt to understand better the mathematical aspect of EMD algorithm \cite{Rilling2003EMD,Wu2004EMD,Flandrin2004EMDa,Rilling2008EMD,Wang2014PA,Huang2009AADA}.  For instance, \citet{Flandrin2004EMDa} found that the EMD algorithm acts as a data-driven wavelet-like expansions. \citet{Wang2014PA} reported that both the time and   space complexity of the EMD algorithm   are  $\mathcal{O}(n\cdot \log n)$, in which $n$ is the data size, but with a larger factor than the traditional Fourier transform.

 \subsection{Hilbert Spectral Analysis}
 With the achieved IMF modes, the Hilbert spectral analysis is then applied to each $C_i(x)$ to retrieve the spectral information via the classical Hilbert transform,  
\begin{equation}
\overline{C}_i(x)=\frac{1}{\pi}P\int \frac{C_i(x')}{x-x'} d x', \label{eq:Hilbert}
\end{equation} 
in which $P$ means the Cauchy principal value. An analytical signal is then reconstructed as, 
\begin{equation}
C_i^A(x)=C_i(x)+j\overline{C}_i(x)=\mathcal{A}_i(x)\exp(j\phi_i(x)),\label{eq:analytical}
\end{equation}
in which $j=\sqrt{-1}$,   $\mathcal{A}_i(x)$ is the amplitude, and $\phi_i(x)$ is the phase function, which are respectively defined as, 
\begin{equation}
\mathcal{A}_i(x)=\vert C_i^A(x)\vert=\sqrt{C_i(x)^2+\overline{C}_i(x)^2},
\end{equation}
for the amplitude, and
\begin{equation}
\phi_i(x)=\arctan\left( \frac{\overline{C}_i(x)}{C_i(x)}\right),
\end{equation}
for the phase function. 
 An instantaneous wavenumber is then defined as, 
 \begin{equation}
 k_i(x)=\frac{1}{2\pi}\frac{d\phi_i(x)}{d x}\label{eq:wavenumber}
 \end{equation}
 \red{Note that the EMD decomposes the given signal very locally into several IMF modes, and the above described HSA approach extracts the instantaneous amplitude $\mathcal{A}_i(x)$ and wavenumber $k_i(x)$ also at a very local level. The EMD-HSA approach thus inherits a very local ability, namely the amplitude- and frequency/wavenumber-modulation to characterize the nonlinear and nonstationary properties of the data collected from the real world
 \citep{Huang1998EMD,Huang1999EMD,Schmitt2016Book}. }

To show the capability of the EMD-HSA approach, we consider here a toy model with two components on the range $-10\le x\le 10$, 
\begin{equation}
z(x)=z_1(x)+z_2(x),\,z_1(x)=\sin(x^2),\,\,z_2(x)=x^2/20,
\end{equation}
The first component $z_1(x)$ has an instantaneous wavenumber $k(x)={\vert(x)\vert}/{2\pi}$.
After the EMD, one IMF mode $C_1(x)$ with one residual $r_1(x)$ are obtained. 
Figure \ref{fig:example} shows a) the toy model $z(x)$, and b)  $C_1(x)$, $r_1(x)$ (thin lines), $z_1(x)$ and $z_2(x)$, respectively. Visually, except  for the range $-2<x<2$, two components are well separated by the EMD algorithm. The instantaneous wavenumber $k(x)$ is retrieved by applying equations \eqref{eq:Hilbert}$\sim $\eqref{eq:wavenumber}. Note that the estimated $k(x)$ agrees with the theoretical one very well, showing the  very local capability of the EMD-HSA approach.

 \subsection{Hilbert-based High-Order Statistics}
One can construct  pairs of the instantaneous wavenumber and amplitude, i.e.,  $[k_i(x), \mathcal{A}_i(x)]$ for all IMF modes. 
A joint probability density function (pdf) $p(k,\mathcal{A})$ is then extracted from all IMF modes \citep{Huang2008EPL,Schmitt2016Book}.
A $k$-condition $q$th-order statistics is defined as, 
\begin{equation}
\mathcal{L}_q(k)=\left\langle \sum_i \mathcal{A}_i^q(x) \vert k_i(x)=k\right\rangle_{x,t},
\end{equation}
where $<\,\cdot\,>_{x,t}$ means an ensemble average over space and time.
In case of scale invariance, one has power-law behavior, 
\begin{equation}
\mathcal{L}_q(k)\sim k^{-\zeta(q)},
\end{equation}
in which $\zeta(q)$ is the Hilbert-based scaling exponent. 
For a simple scaling process, such as fractional Brownian motion, the measured $\zeta(q)$ is equivalent to the one provided by the structure function analysis \citep{Huang2008EPL,Huang2013PRE,Schmitt2016Book}. For a real data with energetic structures,  
 this approach has a capability to isolate those structures to reveal more accurate scaling behavior
 \citep{Huang2010PRE,Huang2013PRE,Tan2014PoF,Schmitt2016Book}.
For more details about the EMD-HSA method, we refer to  Refs. \citep{Huang1998EMD,Huang1999EMD,Schmitt2016Book}.

\section{Results}\label{sec:result}

\begin{figure}
\centering
\includegraphics[width=0.95\linewidth,clip]{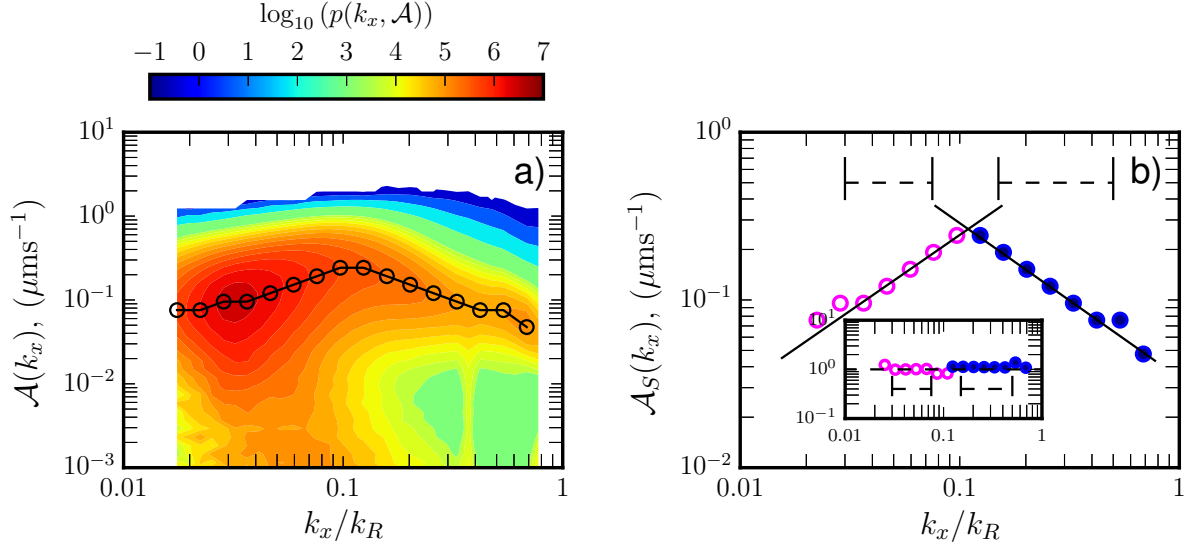}
\caption{(Color online)  a) Measured joint-pdf $p(k,\mathcal{A})$ of the wavenumber $k$ and amplitude $\mathcal{A}$. A skeleton defined by equation \ref{eq:skeleton} is illustrated  by $\ocircle$, showing a scaling trend. The horizontal axis is normalized by the  wavenumber of the bacterial length, i.e., $k_R$. b)  Reproduce the measured skeleton $\mathcal{A}_S(k)$ of the joint-pdf $p(k,\mathcal{A})$. A dual-power-law behavior $\mathcal{A}_S(k)\sim k^{-\gamma}$ is visible with  scaling
exponents $-0.95\pm0.02$   and $0.95\pm0.02$ on the range 
$0.15<k/k_{R}<0.5$ 
%$0.03<k<0.1$  
for the small-scale structures and
 %$0.006<k<0.015$
  $0.03<k/k_{R}<0.075$ for the large-scale structures. The inset shows the compensated
curve to emphasize the observed power-law behavior. }
\label{fig:JPDF}
\end{figure}

\begin{figure}
\centering
\includegraphics[width=0.95\linewidth,clip]{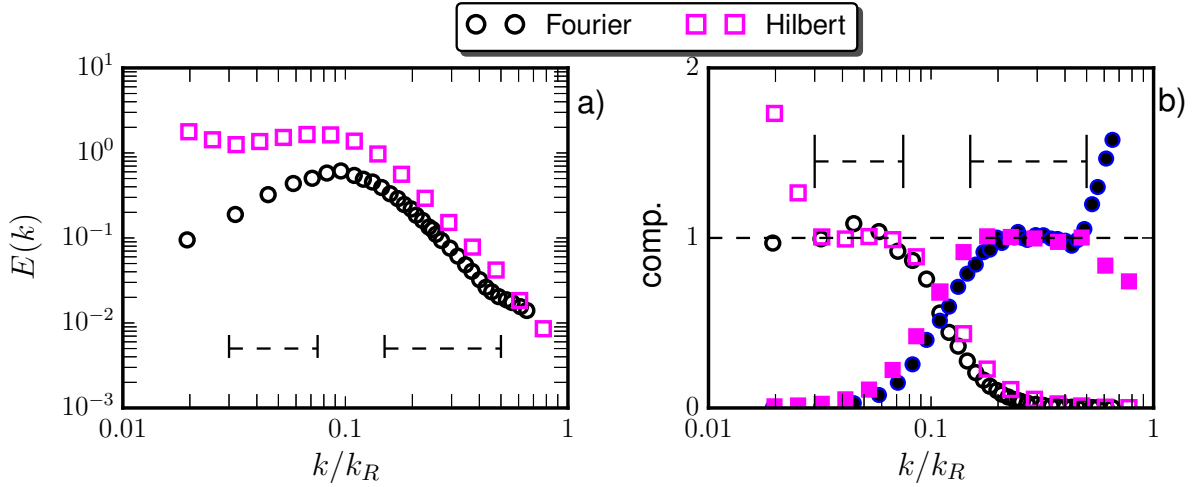}
\caption{(Color online) a) Comparison of the experimental energy spectra $E(k)$ provided by the Fourier analysis ($\ocircle$) and Hilbert spectral analysis ($\square$).  For display convenience, the curve has been vertical shifted. b) The compensated curves using the fitted scaling exponents respectively $\beta_S^{F}=2.68\pm0.06$, $\beta_L^{F}=-1.33\pm0.20$ provided by the Fourier spectrum, and $\beta_S^{H}=2.64\pm0.04$, $\beta_L^{H}=-0.41\pm0.05$ provided by the Hilbert spectrum. The power-law range predicted by the Hilbert approach is indicated by the horizontal dashed line  for the range $0.15<k/k_{R}<0.5$ of the small-scale structures and
  $0.03<k/k_{R}<0.075$ of the large-scale structures.}
\label{fig:spectrum}
\end{figure}

\begin{figure}
\centering
\includegraphics[width=0.95\linewidth,clip]{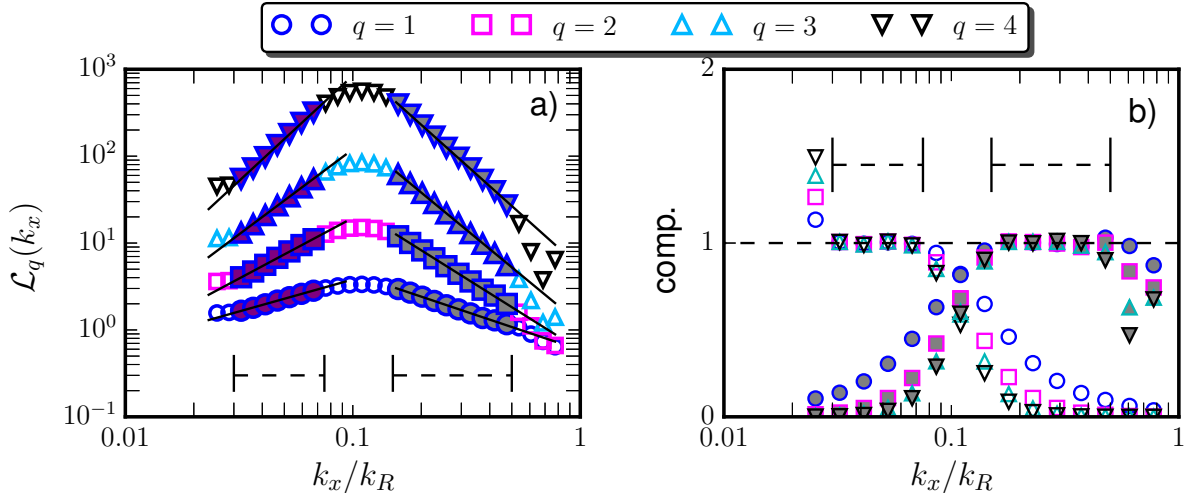}
  \caption{(Color online) a) Measured $q$th-order Hilbert moment $\mathcal{L}_q(k)$.  b) The corresponding compensated curve using the fitted scaling exponent and prefactor. A double power-law behavior is observed on the range 
  $0.03<k/k_{R}<0.075$   and  $0.15<k/k_{R}<0.5$. The existence of the plateau confirms the observed power-law behavior. The scaling exponent is then estimated on these ranges using a least square fitting algorithm.}\label{fig:AHS}
\end{figure}

\begin{figure}
\centering
\includegraphics[width=0.95\linewidth,clip]{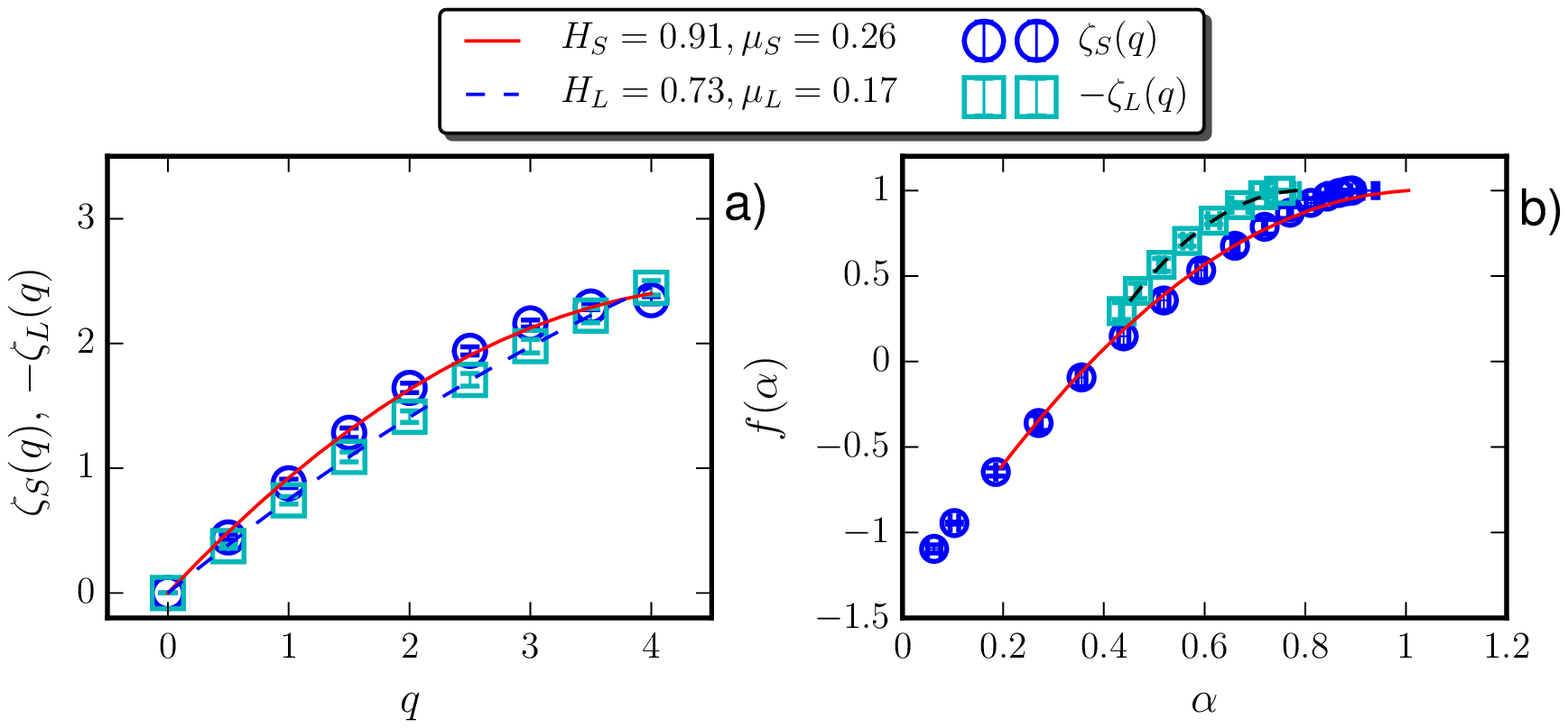}
  \caption{(Color online) a) Experimental scaling exponent $\zeta(q)$  for the small-scale scaling ($\ocircle$) and large-scale scaling ($\square$).  A lognormal formula fitting is also shown as solid and dashed lines respectively for the small and large scales. b) The corresponding  singularity spectrum $f(\alpha)$ versus $\alpha$. The errorbar indicates the  $95\%$ confidence interval provided by the least square fitting algorithm.}\label{fig:scaling}
\end{figure}

\red{In the following the analysis is done along $x$ direction by dividing the Eulerian velocity $\mathbf{u}(x,y)$ into $84$ lines. The EMD-HSA approach is then performed to each slice and the statistics are then averaged over these $84$ lines and all snapshots.}

Figure \ref{fig:JPDF}\,a) shows the measured joint-pdf $p(k,\mathcal{A})$, in which the horizontal axis is normalized by the wavenumber $k_{R}$ of the bacterial body length. For display convenience, the measured $p(k,\mathcal{A})$ has been represented in log scale. A DPL trend is visible  respectively on the range $0.15<k/k_{R}<0.5$ for the small-scale structures, and $0.03<k/k_{R}<0.075$ for the large-scale structures. The scaling trend is characterized by  a skeleton, which is defined as, 
\begin{equation}
p_{\max}(k)=p\left(k,\mathcal{A}_S(k)\right)=\max_{\mathcal{A}}\left\{ p(k,\mathcal{A})\vert_{k} \right\},\label{eq:skeleton}
\end{equation}
The measured $\mathcal{A}_S(k)$ is reproduced in Figure \ref{fig:JPDF}\,b).  The  DPL behavior  is identified, 
\begin{equation}
\mathcal{A}_S(k)\sim k^{-\gamma},
\end{equation}
in which $\gamma$ is the scaling exponent.  Figure \ref{fig:JPDF}\,b) reproduces the measured $\mathcal{A}_S(k)$, showing the DPL behavior. The experimental scaling exponents are respectively $\gamma_{S}=0.95\pm0.02$ for the small-scale structures,  and $\gamma_{L}=-0.95\pm0.02$ for the large-scale structures. 
To emphasize the observed power-law behavior, the compensated curve is shown as the inset in Figure \ref{fig:JPDF}\,b). A clear plateau confirms the existence of the power-law behavior.  \red{The peak location (resp. the viscosity wavenumber $k_{\nu}$) in Fig.\ref{fig:JPDF}\,b)} is to be around $k_{\nu}/k_R\simeq 0.1$, which agrees very well with the observation of the Fourier power spectrum \citep{Wensink2012PNAS}, see also Fig.\,\ref{fig:spectrum}\,a).

Figure \ref{fig:spectrum}\,a)  shows  the measured energy spectrum provided by the Fourier analysis ($\ocircle$) and the Hilbert spectral analysis ($\square$).  The DPL predicted by the Hilbert spectrum is indicated by the horizontal dashed line respectively on the range $0.03<k/k_{R}<0.075$ and $0.15<k/k_{R}<0.5$. To emphasize the observed DPL, the compensated curve, e.g., $E(k)k^{\beta}C^{-1}$, using the fitted scaling exponent $\beta$ and the prefactor $C$ is shown in Fig.\,\ref{fig:spectrum}\,b). The fitted scaling exponents are $\beta^F_S=2.68\pm0.06$, $\beta^F_{L}=-1.33\pm0.20$ provided by the Fourier spectrum, and  $\beta^H_S=2.64\pm0.05$, $\beta^H_{L}=-0.41\pm0.05$ provided by the Hilbert spectrum, respectively. The observed plateau in Fig.\,\ref{fig:spectrum}\,b) confirms again the existence of the DPL behavior at least for the second-order statistics. The statistics of the small-scale fluctuations (resp. the high wavenumber part) by the Fourier and Hilbert agree well with each other. However,  the ones of the large-scale fluctuations (resp. low wavenumber part) do not agree. One possible reason might be the nonlinear distortion embedded in the data \citep{Huang1998EMD}.  Moreover, the DPL is separated by a peak around $k_{\nu}/k_R\simeq 0.1$, which corresponds to the scale of the   fluid viscosity.
 The observed power-law range is limited due to the constrain of this system, e.g., injection scale $R$, the fluid viscosity $k_{\nu}$, the measurement area $L$, etc. 
 
Note that the power-law behavior of the measured spectrum often indicates a cascade process. Analogy  to the 2D turbulence theory, we speculate that at least the energy transfers from the injected scale $R$ to larger scale structures via an inverse cascade.  As mentioned above, due to the fluid viscosity, the energy is then accumulated around $k/k_R\simeq0.1$.  This postulation should be verified carefully via a scale-to-scale energy or enstrophy flux \cite{Zhou2015JFM}. Below we check the high-order statistics to see potential intermittent correction.

Figure \ref{fig:AHS} shows the measured high-order Hilbert moments $\mathcal{L}_q(k)$ for $0\le q\le 4$. The DPL behavior is observed for all $q$ considered here. The power-law ranges are the same as the ones observed in Figure \ref{fig:spectrum}. The corresponding scaling exponents are then estimated using a least-square fitting algorithm. The measured $\zeta(q)$  are shown in Figure \ref{fig:scaling}\,a).  The errorbar indicates the $95\%$ confidence interval provided by the fitting algorithm.
   Visually, the experimental scaling exponent curves are convex, implying  multifractal nature of this active system. To characterize the intensity of multifractality quantitatively, we introduce here a lognormal formula to fit the observed scaling exponent, 
      \begin{equation}
   \zeta(q)=qH-\frac{\mu}{2}\left( q^2H^2-qH\right),\label{eq:lognormal}
   \end{equation}
where $H$ is the Hurst number, and $\mu$ is the intermittency parameter \citep{Li2014PhysicaA}. 
Note that the lognormal model is firstly introduced by \citet{Kolmogorov1962} in 1962 for the Eulerian velocity by assuming a lognormal distribution of the energy dissipation field. It yields for the turbulent velocity $\zeta(q)=q/3-\mu/2(q^2/9-q/3)$ \cite{Frisch1995,Schmitt2016Book}.
   \red{For a given $H$, the intermittency parameter $\mu$ characterizes the deviation from the linear relation $qH$. Or in other words, a larger value of $\mu$ has, the more intermittent the field is. The measured Hurst number and intermittency parameter are $H_S=0.91\pm0.02$ and $\mu_S=0.26\pm0.01$ for $\zeta_S(q)$, and $H_L=0.73\pm0.01$ and $\mu_L=0.17\pm0.01$ for $-\zeta_L(q)$, respectively. It shows a more intermittent small-scale fluctuations.} 

\section{Discussions}\label{sec:discussion}
There are two free parameters in equation \eqref{eq:lognormal}. Therefore, a different choice of $H$ could lead to a different estimated intermittent parameter $\mu$. To avoid this difficulty, we consider below the singularity spectrum  $f(\alpha)$  via the Legendre transform, 
\begin{equation}
\alpha=\frac{d \zeta(q)}{d q},\,f(\alpha)=\min_{q}\left\{ \alpha q-\zeta(q)+1  \right\},
\end{equation} 
in which $\alpha$ is known as the generalized Hurst number or intensity of multifractality \citep{Frisch1995}.  Generally, the broader measured  $\alpha$ and $f(\alpha)$ are the more the experiment $\zeta(q)$ deviates from a linear relation $qH$ even the Hurst number $H$ cannot be accessed precisely.  Thus the analyzed field  is more intermittent
\cite{Frisch1995}.    Figure \ref{fig:scaling}\,b) shows the measured $f(\alpha)$ versus $\alpha$. A broad range of $\alpha$ and $f(\alpha)$  is observed, suggesting that  both small-scale and large-scale fluctuations possessing intermittent correction, while the former one is more intermittent than the latter one, which confirms the result of   the lognormal formula fitting.

We would like to provide some comments on the finite scaling range detected by the Hilbert method. In this special dynamic system, the scaling range is determined by several parameters. They are, at least, the bacterial body length $R\simeq5\,\mu$m, where the energy is injected into the system;  the size of the microfluidic device or the measurement area $L\times L$ with $L\simeq217\,\mu$m for the current data set; the  fluid viscosity
scale $\ell_{\nu}\simeq 50\,\mu$m, \red{below which} a part of the kinetic energy might be dissipated
into heat; the Ekman-like friction provided by interface between the  fluid and the bottom of the microfluidic device and other unknown mechanisms, in which the energy is damped, etc. Note that the fluid viscosity  could be also a function of species and concentrations of bacteria \citep{Sokolov2009PRL,Rafai2010PRL,Lopez2015PRL}. The scaling range of such bacterial turbulence is thus limited due to these  length scales. For instance, the scaling ranges identified in this work are respectively $0.03<k/k_R<0.075$ and $0.15<k/k_R<0.5$, corresponding to  roughly $\simeq 0.4$ and $\simeq 0.5$ decades.  For the former scaling range, it could be limited by the size of the microfluidic device  and the fluid viscosity, i.e., $\ell_{\nu}\simeq10R$ or $k_{\nu}\simeq 0.1 k_R$. It thus could be extended by increasing the measurement area. The latter one is constrained  not only by the fluid viscosity, but also by the bacterial body length $R$ and \red{the depth of the fluid $H_c$. For the spatial scale comparable with the fluid depth $H_c$, the motion could exhibit  3D statistics.}   
It seems that it is difficult to extend this scaling range by simply increasing the measurement resolution or reducing the bacterial body length $R$  since the fluid viscosity is a function of bacterial concentrations and other conditions \citep{Sokolov2009PRL,Rafai2010PRL,Lopez2015PRL}.  

 Moreover, the observed DPL is on the left side of the injection scale. It is therefore then inverse cascade, at least in the sense of the kinetic energy.  In the view of the traditional 2D turbulence, the inverse energy cascade is found to be nonintermittent \citep{Boffetta2012ARFM}. The corresponding forward enstrophy cascade is intermittent if the Ekman friction is present \citep{Nam2000PRL}, which has been confirmed for both the vorticity field \citep{Tan2014PoF} and the velocity field \citep{Wang2015JSTAT}. The Ekman-Navier-Stokes equation for the classical 2D turbulence is written as, 
\begin{equation}\label{eq:NSE}
\partial_t \mathbf{u}+\mathbf{u} \cdot\nabla \mathbf{u}=-\nabla p +\nu \nabla^2
\mathbf{u} -\xi \mathbf{u}+\mathbf{f}_{u},
\end{equation}
in which $\xi$ stands for the Enkman friction coefficient, and $\mathbf{f}_{u}$ is the external forcing, where the energy and enstrophy are injected into the system.  \red{Note that the Ekman friction is a linear drag to model  the three-dimension of no-slip boundary condition or the effect
of the boundary layer  itself in the two-dimensional
description.}
The dual-cascade theory proposed by Kraichnan has been proved partially by the experiments and numerical simulations \citep{Boffetta2012ARFM}.
A continuum model has been put forward to model the bacterial turbulence, which is written as, 
\begin{widetext}
\begin{equation}\label{eq:CT}
\partial_t \mathbf{u}+ \lambda_0\mathbf{u} \cdot\nabla \mathbf{u}=-\nabla p +\Gamma_0\nabla^2\mathbf{u}+\lambda_1 \nabla
\mathbf{u} ^2-(\varpi+ \chi\vert \mathbf{u}\vert^2) \mathbf{u}-\Gamma_2(\nabla^2)^2\mathbf{u}.
\end{equation}
\end{widetext}
where $p$ denotes pressure, and $\lambda_0>1$; $\lambda_1>0$ for the pusher-swimmers as used in this study; $(\varpi,\chi)$   corresponds to a quartic Landau-type velocity potential; $(\Gamma_0,\Gamma_2)$ provides the description of the self-sustained mesoscale turbulence in incompressible active flow, e.g., $\Gamma_0<0$ and $\Gamma_2>0$, the model results in a turbulent state \citep{Wensink2012PNAS}. 
 Comparing the  r.h.s.  of  equations (\ref{eq:CT}) and (\ref{eq:NSE}), \red{one can find that in the continuum theory several additional nonlinear interaction terms are introduced.} We speculate here that \red{these additional nonlinear interactions trigger the intermittency effect into the inverse cascade} of the bacterial turbulence, \red{which is different with the traditional two-dimensional  turbulence and deserves a further careful investigation by checking the scale-to-scale energy/enstrophy flux of this active system.}

\section{Conclusion}

\red{In summary, in this paper the experimental Eulerian velocity of the bacterial turbulence provided by Professor Goldstein at Cambridge University UK was analyzed to emphasize on the multiscaling property. A kind of bacteria  \textit{B subtilis} with a body size $5\mathrm{\mu m}$ is used in this experiment with a volume filling fraction  $84\%$ and a finite depth $\simeq 5\mathrm{\mu m}$. With these parameters, the active flow is in the turbulent phase.
Due to the scale mixture problem, the conventional structure function analysis fails to detect the 
power-law behavior.  A Hilbert-based method was then performed in this work to identify  the scaling behavior.}
A dual-power-law behavior separated by the viscosity wavenumber $k_{\nu}$ is observed with a limit scaling range, which is the result of this special system. This DPL belongs to the inverse cascade since it is on the left side of the injection scale, i.e., \red{$k<k_R$, $k_R$ is the body size wavenumber. As mentioned above for the traditional two-dimensional turbulence, there is no intermittent correction in the inverses cascade. 
On the contrary, due to several additional  nonlinear interactions in this bacterial turbulence, the DPL is found experimentally to be intermittent.} The intensity of the intermittency or multifractality is then characterized by a lognormal formula with measured  $H_S=0.91$ and $\mu_S=0.26$ for the small-scale \red{(resp. high wavenumber part $0.15<k/k_R<0.5$)} fluctuations, and $H_L=0.73$ and $\mu_L=0.17$ for the large-scale \red{(resp. low wavenumber part $0.03<k/k_R<0.075$)} fluctuations, showing that the former cascade is more intermittent than the latter one. This is also confirmed by the calculated singularity spectrum $f(\alpha)$. \red{When comparing a continuum model of this active fluid system with the traditional two-dimensional Ekman-Navier-Stokes equation, there exist several additional nonlinear interactions that 
trigger the intermittentcy in the inverse cascade.}  A less intermittent large-scale fluctuation could be an effect of the fluid viscosity since it plays an important role when $k/k_R\le 0.1$.  We emphasize here that the observed DPL could not be universal since the bacterial turbulence depends on many different parameters, such as, the species of the bacteria, the concentration, etc. It should be studied systematically by applying this Hilbert-based approach.

\begin{acknowledgments}
We  acknowledge the anonymous referees for their useful suggestions.
This work is partially sponsored by the National Natural Science Foundation of China under Grant (No. 11202122, 11222222, 11572185  and 11332006), and partially by the Fundamental Research Funds for the Central Universities (Grant No. 20720150069 (Y.H.), 20720150075 (M.C.)). Y.X. is also supported partially by the Sino-French (NSFC-CNRS) joint research project (No. 1151101101). 
We thank Prof. R.E. Goldstein  for  providing us the experiment data, which can be found at \href{http://damtp.cam.ac.uk/user/gold/datarequests.html}{http://damtp.cam.ac.uk/user/gold/datarequests.html}.
 Y.H. thanks  Dr. G. Rilling and Prof. P. Flandrin from laboratoire de Physique, CNRS \& ENS
Lyon (France) for sharing their Empirical Mode Decomposition (EMD) \textsc{Matlab} codes, which is available at:
{{ http://perso.ens-lyon.fr/patrick.flandrin/emd.html}}. A source package to realize the Hilbert spectral analysis  is available at: {https://github.com/lanlankai}.
\end{acknowledgments}

%\newpage
 %\bibliographystyle{apsrev}
%\bibliographystyle{aipnum4-1}
%\bibliography{all}% Produces the bibliography via BibTeX.

%merlin.mbs apsrev4-1.bst 2010-07-25 4.21a (PWD, AO, DPC) hacked
%Control: key (0)
%Control: author (8) initials jnrlst
%Control: editor formatted (1) identically to author
%Control: production of article title (-1) disabled
%Control: page (0) single
%Control: year (1) truncated
%Control: production of eprint (0) enabled
%

\end{document}